\documentclass[ amsmath, amssymb, aps, prl, twocolumn, lengthcheck, sort&compress, showpacs ]{revtex4-1}    

\usepackage{graphicx}
\usepackage{bm}
\usepackage{dsfont}
\usepackage{units}

\newcommand{\vect}[1]{{\mbox{\boldmath $#1$}}}

\def\Slash#1{\setbox0=\hbox{$#1$} 
\dimen0=\wd0 
\setbox1=\hbox{/} \dimen1=\wd1 
\ifdim\dimen0>\dimen1 
\rlap{\hbox to \dimen0{\hfil/\hfil}} 
#1 
\else 
\rlap{\hbox to \dimen1{\hfil$#1$\hfil}} 
/ 
\fi}

\begin{document}

\title{Nucleon mass from a covariant three-quark Faddeev equation}

\author{G.~Eichmann$^{1,2}$, R.~Alkofer$^2$, A.~Krassnigg$^2$,  D.~Nicmorus$^2$}

\affiliation{$^1$ Institute for Nuclear Physics, Darmstadt University of Technology, 64289 Darmstadt, Germany  \\
             $^2$ Institut f\"ur Physik, Karl-Franzens-Universit\"at Graz, 8010 Graz, Austria}

\date{\today}

\begin{abstract}
            We report the first study of the nucleon where the full Poincar\'e-covariant structure of the three-quark
            amplitude is implemented in the Faddeev equation. We employ an interaction kernel which is consistent with
            contemporary studies of meson properties and aspects of chiral symmetry and its dynamical breaking, thus yielding
            a comprehensive approach to hadron physics. The resulting current-mass evolution of the nucleon mass
            compares well with lattice data and deviates only by $\sim 5 \%$ from the quark-diquark result obtained
            in previous studies.
\end{abstract}

\pacs{%
11.10.St, 
12.38.Lg, 
14.20.Dh 
}
\keywords{}

\maketitle


            Experiments, and hereby especially electroweak probes at all energy
            scales, have provided detailed information about the structure of the
            nucleon. Nevertheless, understanding the nucleon's structure in terms of
            quarks and gluons, the elementary degrees of freedom of Quantum
            Chromodynamics (QCD), has remained a challenge in theoretical hadron physics.

            Starting with the original work of Faddeev \cite{Faddeev:1960su} a formalism has been
            developed to treat a relativistic three-body problem \cite{Taylor:1966zza,Boehm:1976ya,Loring:2001kv}.
            In its covariant form, the Faddeev equation is the three-body analogue of the two-body
            Bethe-Salpeter equation (BSE) \cite{Salpeter:1951sz}. In the case of the nucleon, its
            solution is a covariant three-quark amplitude whose relativistic spin structure has been
            explored in \cite{Machida:1974xw,Henriques:1975uh} and in the light-front formalism
            in \cite{Weber:1986qw,Beyer:1998xy,Karmanov:1998jp,Sun:2001ir}.
            A complete classification according to the Lorentz group and the permutation group $\mathbb{S}_3$
            was derived in \cite{Carimalo:1992ia} in terms of covariant three-spinors.

            The formalism of QCD's Dyson-Schwinger equations (for recent reviews, see e.g.~\cite{Fischer:2006ub,Roberts:2007jh})
            provides a way to embed the covariant three-quark Faddeev equation in a consistent quantum-field theoretical setup.
            The dynamical ingredients in the equation (the dressed quark propagator and the three-quark kernel) can then
            be treated in perfect correspondence with studies of quark and meson properties as well as related aspects of QCD.

            The biggest obstacle on the way to a direct numerical solution of the three-body bound-state equation
            is its complexity. Simplifications employed in the past implemented perturbative quark propagators
            \cite{Kielanowski:1979eb,Falkensteiner:1981ab}, together with a three-body spectator approximation \cite{Stadler:1997iu}, or in a
            Salpeter-equation setup with instantaneous forces \cite{Loring:2001kv}. The corresponding equation of a scalar
            three-particle system with scalar two-body exchange was recently investigated and compared to the light-front
            approach \cite{Karmanov:2008bx}.
            Another kind of simplification can be achieved by considering diquark correlations (see e.g.~\cite{Anselmino:1992vg}
            for an overview). While maintaining full Poincar\'e covariance, the quark-diquark model traces the
            nucleon's binding to colored scalar- and axialvector diquarks, thereby simplifying the Faddeev equation to a quark-diquark
            BSE. This strategy has been applied to study nucleon and $\Delta$ properties
            \cite{Hellstern:1997pg,Oettel:2000jj,Eichmann:2008ef,Nicmorus:2008vb}.

            Here we report the first fully Poincar\'e-covariant computation of the nucleon's Faddeev amplitude beyond
            the quark-diquark approximation. The numerical solution of the Faddeev equation is performed after
            truncating the interaction kernel to a ladder dressed-gluon exchange between any two quarks,
            thereby enabling a direct comparison with corresponding meson studies
            as well as earlier investigations of baryons in the quark-diquark model.

            \begin{figure*}[tbp]
                    \begin{center}
                    \includegraphics[scale=0.45]{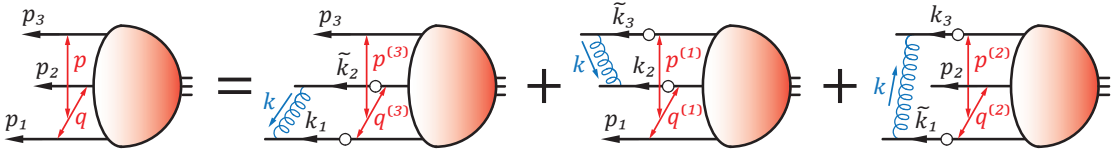}
                    \caption{(color online) Covariant Faddeev equation \eqref{faddeev:eq} in ladder truncation.}\label{fig:faddeev}
                    \end{center}
            \end{figure*}


            In QCD baryons appear as poles in the three-quark scattering matrix.
            This allows one to derive a relativistic three-body bound-state equation:
            \begin{equation}
                \Psi = \widetilde{K}_\text{(3)}\,\Psi\,, \qquad \widetilde{K}_{(3)} = \widetilde{K}_\text{(3)}^\text{irr} + \sum_{a=1}^3 \widetilde{K}^{(a)}_{(2)}\,.
            \end{equation}
            Here, $\Psi$ is the bound-state amplitude defined on the baryon's mass shell. 
            The three-body kernel $\widetilde{K}_{(3)}$
            comprises a three-quark irreducible contribution and the sum of permuted two-quark kernels
            whose quark-antiquark analogues appear in a meson BSE, and
            the superscript $a$ denotes the respective accompanying spectator quark.

            The observation of a strong attraction in the $SU(3)_C$ antitriplet $qq$ channel
            has been the guiding idea for the quark-diquark model, namely that quark-quark correlations
            provide important binding structure in baryons.
            This motivates the omission of the three-body irreducible contribution from the full three-quark kernel.
             The resulting covariant Faddeev equation includes a sum of permuted $qq$ kernels (cf.~Fig.\,\ref{fig:faddeev}):
                \begin{equation}\label{faddeev:eq}
                \begin{split}
                    \Psi_{\alpha\beta\gamma\delta}&(p,q,P) = \\
                      =\sum_{a=1}^3 & \int\limits_k    \widetilde{K}_{\alpha\alpha'\beta\beta'\gamma\gamma'}^{(a)} \, \Psi_{\alpha'\beta'\gamma'\delta}(p^{(a)},q^{(a)},P)\,,
                \end{split}
                \end{equation}
            where $\widetilde{K}^{(a)}$ denotes the renormalization-group invariant products of a $qq$ kernel and two dressed quark propagators:
            \begin{equation} \label{KSS}
                \widetilde{K}_{\alpha\alpha'\beta\beta'\gamma\gamma'}^{(a)} = \delta_{\alpha\alpha'}\mathcal{K}_{\beta\beta''\gamma\gamma''}\, S_{\beta''\beta'}(k_b)  \, S_{\gamma''\gamma'} (\widetilde{k}_c)\,.
            \end{equation}
            $\{a,b,c\}$ is an even permutation of $\{1,2,3\}$ and linked to the respective Dirac index pairs.

            The spin-momentum part of the full Poincar\'e-covariant nucleon amplitude $\Psi_{\alpha\beta\gamma\delta}(p,q,P)$
            is a spin-$\nicefrac{1}{2}$ four-point function
            with positive parity and positive energy:
            it carries three spinor indices $\{\alpha,\beta,\gamma\}$ for the involved valence quarks
            and one index $\delta$ for the spin-$1/2$ nucleon.
            The amplitude depends on the total momentum $P$ and two relative Jacobi momenta $p$ and $q$, where $P^2=-M^2$ is fixed.
            It can be decomposed into 64 Dirac structures:
                \begin{equation}\label{faddeev:amp}
                    \Psi_{\alpha\beta\gamma\delta}(p,q,P) = \sum_{k=1}^{64} f_k \,\tau^k_{\alpha\beta\gamma\delta}(p,q,P),
                \end{equation}
            where the amplitude dressing functions $f_k$ depend on the five Lorentz-invariant combinations
            \begin{equation}\label{mom-variables}
                p^2, \;\; q^2,\;\;   z_0=\widehat{p_T}\cdot\widehat{q_T} ,\;\; z_1 = \widehat{p}\cdot\widehat{P} ,\;\;  z_2 = \widehat{q}\cdot\widehat{P}.
            \end{equation}
            Here, a hat denotes a normalized 4-vector and $p_T^\mu = T^{\mu\nu}_P p^\nu$ a transverse projection with $T^{\mu\nu}_P = \delta^{\mu\nu} - \hat{P}^\mu \hat{P}^\nu$.

    \renewcommand{\arraystretch}{1.2}

            A general spinor four-point function which depends on 3 independent momenta involves 128 independent components of positive parity.
             An orthogonal basis $\{\tau^k\}$ for the 64-dimensional subspace of a positive-parity and positive-energy nucleon
             is given by the set 
            \begin{equation} \label{basisSP}
                \left( \begin{array}{c} \mathsf{S}_{ij}^r \\ \mathsf{P}_{ij}^r \end{array}\right) :=
                \left( \begin{array}{c} \!\!\mathds{1}\otimes\mathds{1} \! \\ \! \gamma^5 \otimes \gamma^5 \!\!\end{array}\right) (\Gamma_i   \otimes  \Gamma_j ) \, (\Lambda^r  \gamma_5 C \otimes \Lambda^+),
            \end{equation}
             where $C=\gamma^4 \gamma^2$ is the charge-conjugation matrix,
             $r = \pm$ refers to the positive- and negative-energy projectors $\Lambda^\pm(P) = (\mathds{1}\pm \widehat{\Slash{P}})/2$,
             and the tensor product is understood as $(A\otimes B)_{\alpha\beta\gamma\delta} = A_{\alpha\beta} B_{\gamma\delta}$.
            The relative-momentum dependence of the basis elements is carried by the $\Gamma_i$, $i=1,2,3,4$, defined by
             \begin{equation}\label{FADEEV:Gamma_i}
                \Gamma_i(p,q,P) = \left\{  \mathds{1},\;
                                                       \textstyle\frac{1}{2}\,\displaystyle[ \widehat{\Slash{p}_T}, \widehat{\Slash{q}_{t}} ],\;
                                                       \widehat{\Slash{p}_T},\;
                                                       \widehat{\Slash{q}_{t}}
                                              \right\}\,.
             \end{equation}
             The momenta $\{\widehat{p_T}, \widehat{q_{t}}, \widehat{P}\}$ were conveniently chosen to be orthonormal with respect to the Euclidean metric via
             \begin{equation}
                 p_T^\mu := T^{\mu\nu}_P  \,p^\nu,  \quad
                 q_{t}^\mu := T^{\mu\nu}_{{p_T}} \,T^{\nu\lambda}_{P} \,q^\lambda = T^{\mu\nu}_{{p_T}} \,q_T^\nu\,.
             \end{equation}

    \renewcommand{\arraystretch}{1.0}

             A partial-wave decomposition leads to linear combinations of the $\{\mathsf{S}_{ij}^r,\,\mathsf{P}_{ij}^r\}$
             as eigenstates of quark-spin and orbital angular momentum operators $\vect{S}^2$ and $\vect{L}^2$
             in the nucleon rest frame. The 64 basis covariants (32 each for total quark spin $s=1/2$ and $s=3/2$, respectively)
             can be arranged into sets of 8 $s$-waves ($l=0$), 36 $p$-waves ($l=1$), and 20 $d$-waves ($l=2$).
             For instance, the dominant contributions to the Faddeev amplitude are given by the $s$-waves
             \begin{equation}
             \begin{split}
                 \gamma_5 C \otimes \Lambda^+ &= \sum_{r=\pm} \mathsf{S}_{11}^r\,,\\
                 \gamma^\mu_T C \otimes \gamma^\mu_T \gamma_5 \Lambda^+ &= \sum_{r=\pm} \left( r \,\mathsf{S}_{22}^r + \mathsf{P}_{33}^r + \mathsf{P}_{44}^r \right),
             \end{split}
             \end{equation}
             with $\gamma_T^\mu = T_P^{\mu\nu} \gamma^\nu$.
             In the quark-diquark model, these correspond to scalar-scalar and axialvector-axialvector combinations of diquark and quark-diquark amplitudes
             for either of the three diagrams appearing in the Faddeev equation.

             The basis elements can be expressed in terms of quark three-spinors frequently used in the literature, e.g. Ref.\,\cite{Carimalo:1992ia}.
              In this context the elements $\mathsf{S}_{11}^+ = \Lambda^+ \gamma_5 C \otimes \Lambda^+$ and
              $\mathsf{A}_{11}^+ := \Lambda^+ \gamma^\mu_T C \otimes \gamma^\mu_T \gamma_5 \Lambda^+$ read
    \renewcommand{\arraystretch}{1.2}
              \begin{equation}\label{faddeev:spinor}
              \begin{array}{rl}
                 -\mathsf{S}_{11}^+ \, U^\uparrow &=  (U^\uparrow  U^\downarrow - U^\downarrow  U^\uparrow) \, U^\uparrow\,,  \\
                  \mathsf{A}_{11}^+ \, U^\uparrow &=  (U^\uparrow  U^\downarrow + U^\downarrow  U^\uparrow) \, U^\uparrow - 2 \, U^\uparrow  U^\uparrow  U^\downarrow\,,
              \end{array}
              \end{equation}
              where the $U^\sigma(P)$ are eigenspinors of $\Lambda^+$ and therefore satisfy the free Dirac equation for a spin-$\nicefrac{1}{2}$ particle.
    \renewcommand{\arraystretch}{1.0}

             The Pauli principle requires the Faddeev amplitude to be antisymmetric under exchange of any two quarks.
             The Faddeev kernel $\widetilde{K}_{(3)}$ is invariant
             under the permutation group $\mathbb{S}_3$. The eigenstates of the Faddeev kernel 
             can hence be arranged into irreducible $\mathbb{S}_3$ multiplets
             \begin{equation}\label{psi:irreps}
                 \Psi_\mathcal{S}, \; \Psi_\mathcal{A}, \; \left(\begin{array}{c} \Psi_\mathcal{M_A} \\ \Psi_\mathcal{M_S} \end{array}\right),
             \end{equation}
             of which the first two (totally symmetric or antisymmetric) solutions are unphysical while the mixed-symmetry doublet constitutes the Dirac part of the nucleon amplitude.
             Taking into account the flavor and color structure, the full Dirac--flavor--color amplitude reads
              \begin{equation}\label{FE:nucleon_amplitude_full-2}
                    \Psi(p,q,P) =  \Big\{ \Psi_\mathcal{M_A}  \mathsf{T}_\mathcal{M_A} + \Psi_\mathcal{M_S}  \mathsf{T}_\mathcal{M_S} \Big\} \frac{\varepsilon_{ABC}}{\sqrt{6}} \,,
              \end{equation}
             where $\mathsf{T}_\mathcal{M_A}$, $\mathsf{T}_\mathcal{M_S}$ denote the isospin-$1/2$ flavor tensors for proton and neutron and
             $\varepsilon_{ABC}$ the antisymmetric color-singlet wave function.
             A flavor-dependent kernel in the Faddeev equation will mix $\Psi_\mathcal{M_A}$ and $\Psi_\mathcal{M_S}$ whose dominant contributions
             are given by $\mathsf{S}_{11}^+$ and $\mathsf{A}_{11}^+$, respectively.
             Similarly to the analogous case of a diquark amplitude, the symmetry does however not
             reduce the number of Dirac covariants since the dressing functions $f_k$ transform under the permutation group as well.


        To proceed with the numerical solution of the Faddeev equation, we need to specify the quark-quark kernel $\mathcal{K}$
        and the dressed quark propagator $S(p)$ which appear in Eq.\,\eqref{KSS}. This is achieved via the
        axial-vector Ward-Takahashi identity which encodes the properties of chiral symmetry in connection with QCD.
        Its satisfaction by the interaction kernels in related equations guarantees the correct implementation of
        chiral symmetry and its dynamical breaking, leading e.g.~to a generalized Gell-Mann--Oakes--Renner relation
        valid for all pseudoscalar mesons and all current-quark masses \cite{Maris:1997hd,Holl:2004fr}.
        In particular the pion, being the Goldstone boson related to dynamical chiral symmetry breaking, becomes massless
        in the chiral limit, independent of the details of the interaction. Specifically, we describe the $qq$ kernel
        by a ladder dressed-gluon exchange:
                \begin{equation}\label{RL:kernel}
                    \mathcal{K}_{\alpha\alpha'\beta\beta'}(k) = Z_2^2 \,\frac{4\pi\alpha(k^2)}{k^2} \, T^{\mu\nu}_k \,\gamma^\mu_{\alpha\alpha'} \,\gamma^\nu_{\beta\beta'}
                \end{equation}
            which must also appear in the corresponding
            quark Dyson-Schwinger equation whose solution defines the renormalized dressed quark propagator:
            \begin{equation}\label{RL:QuarkDSE}
                S^{-1}_{\alpha\beta}(p) = Z_2 \left( i\Slash{p} + m \right)_{\alpha\beta}  + \int_q \mathcal{K}_{\alpha\alpha'\beta'\beta}(k) \,S_{\alpha'\beta'}(q)\,.
            \end{equation}
            The bare quark mass $m$ enters as an input, and the gluon momentum is $k=p-q$.
            The inherent color structure of the kernel leads to prefactors $2/3$ and $4/3$ for the integrals
            in Eqs.\,\eqref{faddeev:eq} and \eqref{RL:QuarkDSE}, respectively.

            Eqs.\,(\ref{RL:kernel}--\ref{RL:QuarkDSE}) define the rainbow-ladder (RL) truncation
            which has been extensively used in Dyson-Schwinger equation studies of mesons and baryons in the quark-diquark model,
            e.\,g.~\cite{Krassnigg:2009zh,Eichmann:2007nn} and references therein.
           The non-perturbative dressing of the gluon propagator and the quark-gluon
	       vertex are absorbed into an effective coupling $\alpha(k^2)$ for which we adopt the ansatz~\cite{Maris:1999nt,Eichmann:2008ae}
        \begin{equation}\label{couplingMT}
            \alpha(k^2) = \pi \eta^7  \left(\frac{k^2}{\Lambda^2}\right)^2 \!\! e^{-\eta^2 \left(\frac{k^2}{\Lambda^2}\right)} + \alpha_\text{UV}(k^2) \,.
        \end{equation}
        The second term reproduces the logarithmic decrease of QCD's perturbative running coupling and vanishes at $k^2=0$.
        The first term is parametrized by an infrared scale $\Lambda$ and a dimensionless parameter $\eta$.
        It yields the non-perturbative enhancement at small and intermediate gluon
	    momenta necessary to generate dynamical chiral symmetry breaking and hence a constituent-quark mass scale.
        ($\{\Lambda,\,\eta\}$ and the infrared parameters used in \cite{Eichmann:2008ae} are related by
        $\mathcal{C}= (\Lambda/\Lambda_t)^3$ and $\omega = \eta^{-1} \Lambda/\Lambda_t$, with $\Lambda_t = 1$ GeV.)

        Beyond the present truncation, corrections arise from pseudoscalar meson-cloud contributions which
        provide a substantial attractive contribution to the `quark core' of dynamically
	    generated hadron observables in the chiral regime and vanish with increasing current-quark mass,
        but also from non-resonant contributions due to the infrared structure of the quark-gluon vertex.
        To anticipate corrections we exploit the freedom in adjusting the input scale $\Lambda$.
        We adopt two different choices established in the literature in the context of $\pi$ and $\rho$
        properties \cite{Eichmann:2008ae}:

        Setup A is determined by a fixed scale $\Lambda = 0.72$ GeV, chosen in \cite{Maris:1999nt} to reproduce
         the experimental pion decay constant and the phenomenological quark condensate. Corresponding results
         are therefore aimed in principle at a comparison to experimental data for meson and baryon properties
         (see \cite{Krassnigg:2009zh,Nicmorus:2008vb} and references therein).
        Setup B defines a current-mass dependent scale which is deliberately inflated close to the chiral limit,
         where $\Lambda \approx 1$ GeV \cite{Eichmann:2008ae}. It is meant to describe a hadronic quark core which
         must subsequently be dressed by pion-cloud effects and other corrections. As a result, $\pi$, $\rho$,
         $N$ and $\Delta$ observables are consistently overestimated, but (with the exception of the $\Delta$-baryon)
         compatible with quark-core estimates from quark models and chiral perturbation theory (for a detailed
         discussion, see \cite{Eichmann:2008ae,Eichmann:2008ef,Nicmorus:2008vb}).
        Irrespective of the choice of $\Lambda$, hadronic ground-state properties have turned out to be insensitive
        to the value of $\eta$ in a certain range \cite{Maris:1999nt,Krassnigg:2009zh}. Consequently, with
        Eqs.~(\ref{RL:kernel}--\ref{couplingMT}) and $\Lambda$, the input of the Faddeev equation is completely specified
        with all parameters already fixed to meson properties.


    \renewcommand{\arraystretch}{1.0}

             \begin{table}[t]

                \begin{center}
                \begin{tabular}{   l @{\;\;} || @{\quad}l@{\quad} || @{\quad}l@{\quad}   |   @{\quad} l      }

                                  &  Q-DQ~\cite{Nicmorus:2008vb}      &      Faddeev ($\mathcal{M_A}$)  &  Faddeev ($\mathcal{M_S}$)     \\   \hline

                    Setup A         &  $0.94$                           &     $0.99$                      &   $0.97$                     \\
                    Setup B         &  $1.26(2)$                        &     $1.33(2)$                   &   $1.31(2)$

                \end{tabular} \caption{Nucleon masses obtained from the Faddeev equation in setups A and B and
                                       compared to the quark-diquark result. The $\eta$ dependence is indicated for
                                       setup B in parentheses.
                                       }\label{tab:results}
                \end{center}

        \end{table}


            \begin{figure}[t]
            \begin{center}
            \includegraphics[scale=0.9]{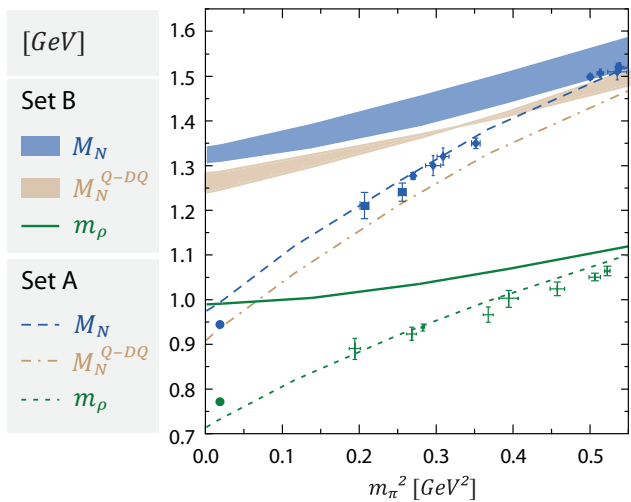}
            \caption{(color online) Evolution with $m_{\pi}^2$ of $m_\rho$ and $M_N$
                                     compared to lattice data; see \cite{Nicmorus:2008vb} for references.
                                     The quark-diquark model result for $M_N$ is plotted for comparison.
                                     Dashed and dashed-dotted lines correspond to setup A;
                                     the solid line for $m_\rho$ and the bands for $M_N$ (mixed-antisymmetric
                                     solution) are the results of setup B, 
                                     where the variation with $\eta$ is explicitly taken into account.
                                     Dots denote the experimental values.} \label{fig:FADDEEV:nucleon-mass}
            \end{center}
            \end{figure}


 \renewcommand{\arraystretch}{1.2}

             Since the dressed-gluon exchange kernel is flavor-independent and we consider only equal quark masses,
             the equations for the Dirac amplitudes $\Psi_\mathcal{M_A}$ and $\Psi_\mathcal{M_S}$ in
             Eq.\,\eqref{FE:nucleon_amplitude_full-2} decouple because of the orthogonality of
             the two flavor tensors $\mathsf{T}_\mathcal{M_A}$ and $\mathsf{T}_\mathcal{M_S}$.
             Hence one obtains two degenerate solutions of the Faddeev equation, where by virtue of the iterative solution method
             the symmetry of the start function determines the symmetry of the resulting amplitude.
             The massive computational demand in solving the equation primarily comes from the
             five Lorentz-invariant momentum combinations of Eq.\,\eqref{mom-variables} upon which the amplitudes depend.
             In analogy to the separability assumption of the nucleon amplitude in the quark-diquark model
             we omit the dependence on the angular variable $z_0=\widehat{p_T}\cdot\widehat{q_T}$
             but solve for all 64 dressing functions $f_k(p^2, q^2,0,z_1,z_2)$.

             The resulting nucleon masses at the physical pion mass in both setups A and B are
             presented in Table~\ref{tab:results}. The difference of $\sim 2\%$ between the
             $\mathcal{M_A}$ and $\mathcal{M_S}$ solutions is presumably an artifact associated with the
             omission of the angle $z_0$. For either solution typically only a small number of covariants are relevant
             which are predominantly $s$-wave with a small $p$-wave admixture. The angular dependence in the
             variable $z_2$ is small compared to $z_1$ in analogy to the quark-diquark model,
             where the dependence on the angle between
             the relative and total momentum of the two quarks in a diquark amplitude is weak.

             The evolution of $M_N$ and the $\rho$-meson mass from the BSE vs.~$m_\pi^2$ is plotted in
	         Fig.~\eqref{fig:FADDEEV:nucleon-mass} and compared to lattice results.
             The findings are qualitatively similar to those for $m_\rho$: setup A, where the coupling strength
             is adjusted to the experimental value of $f_\pi$, agrees with the lattice data, which is reasonable
             in light of a recent study of corrections beyond RL truncation for the $\rho$-meson \cite{Fischer:2009jm}.
             Setup B provides a description of a quark core which overestimates the experimental values while it approaches
             the lattice results at larger quark masses.

             A comparison to the consistently obtained quark-diquark model result exhibits a discrepancy of only $\sim 5\%$.
             This surprising and reassuring result indicates that a description of the nucleon as a superposition of scalar and axial-vector
             diquark correlations that interact with the remaining quark provides a close approximation to the consistent
             three-quark nucleon amplitude.

             We have provided the first fully Poincar\'e-covariant three-quark solution of the nucleon's Faddeev equation.
             The present study contains the first numerical results for the nucleon mass in this approach. Due to the
             considerable computational efforts involved, more results and an in-depth investigation with regard to the complete
             set of invariant variables will follow in subsequent publications.
             Future extensions of the present work will include an analogous investigation of
             the $\Delta$-baryon, more sophisticated interaction kernels, e.g.~in view of pionic corrections,
             and ultimately a comprehensive study of baryon resonances.

                We thank C.\,S.~Fischer, M.~Schwinzerl, and R.~Williams for useful discussions.
                This work was supported by the Helmholtz Young Investigator Grant
                VH-NG-332, the Austrian Science Fund FWF under Projects No. P20592-N16,
                P20496-N16, and Doctoral Program No. W1203, and in part by the European Union
		        (HadronPhysics2 project ``Study of strongly interacting matter'').


\bibliographystyle{apsrev}
\bibliography{had_nucl_graz}

\end{document}